\documentclass[aps,prd,amsmath,amssymb,twocolumn,preprintnumbers,nofootinbib]{revtex4-1}

\pdfoutput=1
\usepackage[utf8]{inputenc} 
\usepackage{graphicx}
 \usepackage{multirow}
\graphicspath{{./figures/}}
\usepackage{url}
\usepackage[bookmarks, pagebackref=false]{hyperref}
\usepackage[usenames,dvipsnames]{xcolor}
\definecolor{orange}{cmyk}{0,0.5,1,0}
\definecolor{rossoCP3}{cmyk}{0,.88,.77,.40}
\definecolor{graa}{rgb}{0.8,0.8,0.8}
\definecolor{blaa}{rgb}{0.2,0.2,0.6}
		\hypersetup{
			colorlinks, 
			bookmarksopen, 
			bookmarksnumbered,
			citecolor=blaa, 		
			linkcolor=rossoCP3,	
			urlcolor=rossoCP3,			
			}
\usepackage{amsthm}
\usepackage{bm}
\usepackage{bbm}
\usepackage{pxfonts}

\usepackage{amsmath,amssymb,amsfonts}
\usepackage{color}
\usepackage{float}
\usepackage{hyperref}
\usepackage[Symbolsmallscale]{upgreek}
\usepackage{amsmath}
\usepackage{amsfonts}
\usepackage{amssymb,dsfont}
\usepackage{graphicx}
\usepackage{amssymb}
\usepackage[vcentermath]{youngtab}
\usepackage[all]{xy}
\usepackage{pstricks}
\usepackage{dsfont}%
\usepackage{placeins}
\usepackage{xspace}
\usepackage{cancel} 

\usepackage{slashed}
\usepackage[caption=false, labelformat=simple, listofformat=subsimple, labelfont=default, margin=5pt, justification=raggedright]{subfig}

	\makeatletter
		\renewcommand{\p@subfigure}{}
	\makeatother

\usepackage{natbib}

\usepackage{feynmf}

\newcommand{\beq}{\begin{eqnarray}}
\newcommand{\eeq}{\end{eqnarray}}

\newcommand{\bmp}{\noindent\begin{minipage}{16cm}}
\newcommand{\emp}{\end{minipage}\vskip 7mm} 

\def\lsim{\mathrel{\rlap{\lower4pt\hbox{\hskip1pt$\sim$}}
    \raise1pt\hbox{$<$}}}                
\def\gsim{\mathrel{\rlap{\lower4pt\hbox{\hskip1pt$\sim$}}
    \raise1pt\hbox{$>$}}}                

\begin{document}


\title{\texorpdfstring{\Large\color{rossoCP3}   Conformal Window 2.0: The Large $N_f$ Safe Story}{Higgs Critical Exponents and Conformal Bootstrap in Four Dimensions}}
\author{Oleg {\sc Antipin}$^{\color{rossoCP3}{\clubsuit}}$}
 \author{Francesco {\sc Sannino} $^{\color{rossoCP3}{\diamondsuit}}$}
\affiliation{\mbox{ $^{\color{rossoCP3}{\clubsuit}}$ Rudjer Boskovic Institute, Division of Theoretical Physics, Bijeni\v cka 54, HR-10000 Zagreb, Croatia}\\\mbox{{ $^{\color{rossoCP3}{\diamondsuit}}$\color{rossoCP3} {CP}$^{ \bf 3}${-Origins}} \&  {\color{rossoCP3}\rm{Danish IAS}},  University of Southern Denmark} \\ $^{\color{rossoCP3}{\diamondsuit}}$CERN, Theoretical Physics Department, Switzerland}

\begin{abstract}
We extend the phase diagram of SU(N) gauge-fermion theories as function of number of flavours and colours to the region in which asymptotic freedom is lost. We argue, using large $N_f$ results, for the existence of an ultraviolet interacting fixed point at sufficiently large number of flavours opening up to a second ultraviolet conformal window in the number of flavours vs colours phase diagram. We first review the state-of-the-art for the large $N_f$ beta function and then estimate the lower boundary of the ultraviolet window. The theories belonging to this new region are examples of  safe non-abelian quantum electro dynamics, termed here {\it safe QCD}.  Therefore, according to Wilson,  they  are fundamental. An important critical quantity is the fermion mass anomalous dimension at the ultraviolet fixed point that we determine at leading order in $1/N_f$. We discover that its value is comfortably below the bootstrap bound.  We also investigate the abelian case and find that at the potential ultraviolet fixed point  the related fermion mass anomalous dimension has a singular behaviour suggesting that a more careful investigation of its ultimate fate is needed. 
\\
[.3cm]
{\footnotesize  \it Preprint: CP$^3$-Origins-2017-033 DNRF90, CERN-TH-2017-187
}
\end{abstract}
\maketitle

The discovery of asymptotic freedom \cite{Gross:1973ju,Politzer:1973fx} has been a landmark in our understanding of fundamental interactions. By fundamental we mean that, following Wilson~\cite{Wilson:1971bg,Wilson:1971dh}, these theories are valid at arbitrary short and long distance scales.    Asymptotic freedom has therefore guided a great deal of Standard Model (SM) extensions.   Likewise the discovery of four dimensional asymptotically safe field theories \cite{Litim:2014uca}  constitutes an important alternative to asymptotic freedom.  It has opened the door to new ways to generalise the Standard Model \cite{Abel:2017ujy,Abel:2017rwl,Pelaggi:2017wzr,Mann:2017wzh,Pelaggi:2017abg,Bond:2017wut} with impact in dark matter physics and cosmology \cite{Pelaggi:2017abg}. In the original construction \cite{Litim:2014uca} elementary scalars and their induced Yukawa interactions played a crucial role in helping make the overall gauge-Yukawa theory safe.  Here we will investigate, instead, the ultraviolet fate of gauge-fermion theories at finite number of colours but very large number of flavours of both abelian and nonabelian nature. 

We start by considering an $SU(N_c)$ gauge theory with $N_f$ fermions transforming according to a given representation of the gauge group. We will assume that asymptotic freedom is lost, meaning that the number of flavours is larger than $N_f^{AF} > 11C_G/(4T_R)$, where the first coefficient of the beta function changes sign. We do not need to specify the fermion representation, but will give explicit examples later.  In any case, for normalisation purposes, we recall that in the fundamental representation the relevant group theory coefficients  are  $C_G=N_c$, $C_R=(N_c^2-1)/2N_c$ and $T_R=1/2$. At the one loop order the theory is simultaneously  free in the infrared  (non-abelian QED) and trivial, meaning that  the only sensible way to take the continuum limit (i.e. sending the Landau pole induced cutoff to infinity) is for the theory to become non-interacting. At two-loops, in a pioneering work, Caswell \cite{Caswell:1974gg} demonstrated that the sign of the second coefficient of the gauge beta function is such that an UV interacting fixed point  (asymptotic safety) cannot arise when the number of flavours is just above the value for which asymptotic freedom is lost.   This observation immediately implies that for gauge-fermion theories triviality can be replaced by safety only above a new critical number of flavours.  To investigate this possibility a logical limit to consider is the large  $N_f$  one at fixed number of colours \cite{PalanquesMestre:1983zy,Gracey:1996he,Holdom:2010qs,Pica:2010xq}. This will be the focus of our work. 

 \underline{\it Non-abelian large-$N_f$ beta function review:} 
Using the conventions of \cite{PalanquesMestre:1983zy,Holdom:2010qs}, the standard beta function reads
 \begin{equation}
\beta(\alpha)\equiv\frac{\partial \ln \alpha}{\partial \ln \mu}=-b_1\frac{\alpha}{\pi}+...,  \qquad \alpha= \frac{g^2}{4\pi} \ ,
\label{e10}\end{equation}
with $g$ the gauge coupling. At large $N_f$ it is conveniently expressed in terms of the normalised coupling   $A\equiv N_f T_R\alpha/\pi$ and expanding in  $1/N_f$ we can write 
\begin{equation}
\frac{3}{2}\frac{\beta(A)}{A}=1+\sum_{i=1}^{\infty}\frac{H_i(A)}{N_f^i} \ ,
\label{e8}\end{equation}
where the leading identity term corresponds to the one loop result and constitutes the zeroth order term in the $1/N_f$ expansion. If the functions $|H_i(A)|$ were finite then in the large $N_f$ limit the zeroth order term would prevail and the Landau pole would be inevitable. This however is not the case. The occurrence of a divergent structure in the $H_i(A)$ functions renders the situation worth investigating. 

According to the large $N_f$ limit each function $H_i(A)$ re-sums an infinite set of Feynman diagrams at the same order in $N_f$ with $A$ kept fixed. Let's make this point explicit for the leading $H_1(A)$ term. The $nth-$loop beta function coefficients $b_n$ for $n \geq 2$ are polynomials in $T_R N_f$ of lowest degree 0 and highest degree $n-1$:
\beq
b_n=\sum_{k=0}^{n-1} b_{n,k} (T_R N_f)^k \ .
\label{pert}
\eeq
The coefficient with the highest power of $T_R N_f$ will be $b_{n,n-1}$ and this is the coefficient contributing to $H_1(A)$ at the $nth-$loop order.  

Now, the $nth-$loop beta function will have an interacting UV fixed point (UVFP) when the following  equation has a physical zero \cite{Pica:2010xq}
\beq
b_1 + \sum_{k=2}^{n} b_k \alpha^{k-1} = 0 \  \ \ \text{where} \ \ b_1=\frac{11C_G}{6}-\frac{2T_R N_f}{3} \ .
\label{BF}
\eeq
This expression simplifies at large $N_f$.  In fact when truncated at a given perturbative order $nmax$ one finds that the highest loop beta function coefficient $b_{nmax}$ contains just the highest power of $(T_R N_f)^{nmax-1}$ multiplied by the coefficient $b_{nmax,nmax-1}$, as it can be seen   from Eq.\ref{pert}. Since this highest power of $(T_R N_f)^{nmax-1}$ dominates in the $N_f\to \infty$ limit and since in this limit $b_1<0$, the criterium for the existence of a UV zero in the $nmax-$loop beta function becomes \cite{Pica:2010xq}: 
\emph{\begin{center} for $N_f\to \infty \ , \ \  $ $\beta(\alpha)$ has an UVFP only  if $b_{nmax,nmax-1}>0$ \ .\end{center}}

In perturbation theory, only the first few coefficients $b_{n,n-1}$ are known but, remarkably, it is possible to resum the perturbative infinite sum to obtain $H_1(A)$. From the results in \cite{PalanquesMestre:1983zy,Gracey:1996he}  
\begin{eqnarray}
H_1(A)&=&-\frac{11}{4}\frac{C_G}{T_R}+\int_0^{A/3} I_1(x)I_2(x)dx,\label{e7}\\
 I_1(x)&=&{\frac {  ( 1+x ) ( 2\,x-1 )^2  ( 2\,x-3 )^2 \sin ( \pi \,x )^{3} \Gamma 
 ( x-1 )^{2}\Gamma  ( -2\,x ) }{ ( x-2 ) \ {\pi }^{3}}} \nonumber \\
I_2(x)&=&\frac{C_{{R}}}{T_R}+\frac{\left( 20-43\,x+32\,{x}^{2}-14\,{x}^
{3}+4\,{x}^{4} \right) }{ 4\left( 2\,x-1 \right)  \left( 2\,x-3 \right) \left( 1-x^2
 \right) }\frac{C_{{G}}}{T_R}  \ .\nonumber
\end{eqnarray}
By inspecting  $I_1(x)$ and $I_2(x)$ one notices that the  $C_G$ term in $I_2$ has a pole in the integrand at $x=1$ ($A=3$). This corresponds to a logarithmic singularity in $H_1(A)$  that will cause the beta function to have a UV zero already to this order in the $1/N_f$ expansion and, by solving the $1+H_1(A)/N_f=0$ condition, this non-trivial UV fixed point occurs at \cite{Litim:2014uca}:
\beq 
A^*=3-\exp \big[-a \frac{N_f}{N_c}+b\big] \ ,
\label{Kstar}
\eeq
where $a=16 T_R$ and $b=18.49-5.26 \ C_R /C_G$. 

Performing a Taylor expansion of the integrand in Eq.\ref{e7} and integrating term-by-term we can obtain the $nth$-loop coefficients $b_{n,n-1}$ and check our criteria above for the existence of the safe fixed point. This procedure was performed in \cite{Shrock:2013cca} up to 18th-loop order where it was also checked that the first 4-loops agree with the known perturbative results. It was found that, even though up to the 12th-loop order the resulting coefficients are scattered between the positive and negative values, starting from the 13th-loop order all $b_{n,n-1}$ are positive for the fundamental, two-index representations and symmetric/antisymmetric rank-3 tensors. This supports the possible  existence of the UV fixed point. These results have been confirmed, extended and employed to build the first realistic asymptotically safe extensions of the SM \cite{Mann:2017wzh,Abel:2017rwl,Pelaggi:2017abg}. 

This concludes our review  of the large $N_f$ beta-function and its use to investigate the UV fate  of nonabelian QED theories.  If these theories are safe we will call them {\it Safe QCD} \cite{Sannino:2015sel}.  We move now to provide a careful investigation and prediction of the safe large $N_f$ quark mass anomalous dimension.
 
\underline{\it Safe large $N_f$ mass anomalous dimension and bootstrap:}
We start by summarising the general expression for  the large $N_f$ mass anomalous dimension \cite{PalanquesMestre:1983zy} :
\beq
&\gamma_m(A) &\equiv-\frac{\partial \ln m}{\partial \ln \mu}= \sum_{i=1}^{\infty}\frac{G_i(A)}{N_f^i} \ ,\\
&G_1(A)=&\frac{C_R}{2 T_R}\frac{A(1-2A/9)\Gamma(4-2A/3)}{\Gamma(1+A/3)[\Gamma(2-A/3)]^2\Gamma(3-A/3)}
 \ .
\label{anom}
\eeq
We immediately note that the first singularity in the expression for $\gamma_m(A)$ appears at $A=15/2$ while the first singularity of the beta function occurs at the smaller value of $A=3$.

Inserting the UVFP value from Eq.\ref{Kstar} into Eq.\ref{anom}  and taking the limit of $N_f\to \infty$ with $N_c$ fixed, we achieve the UV fixed point for $A^*\to 3$ up to exponentially small corrections yielding 
\beq
 \frac{A(1-2A/9)\Gamma(4-2A/3)}{\Gamma(1+A/3)[\Gamma(2-A/3)]^2\Gamma(3-A/3)} \stackrel{A\to 3}\to 1 \ ,
\eeq
implying via Eq.\ref{anom} that
\beq
\gamma_m^*(A) \stackrel{N_f\to \infty}\to \frac{C_R }{2 T_R N_f} \ .  
\label{asymptotic}
\eeq

\begin{figure}[h!]
\includegraphics[width=.5\textwidth]{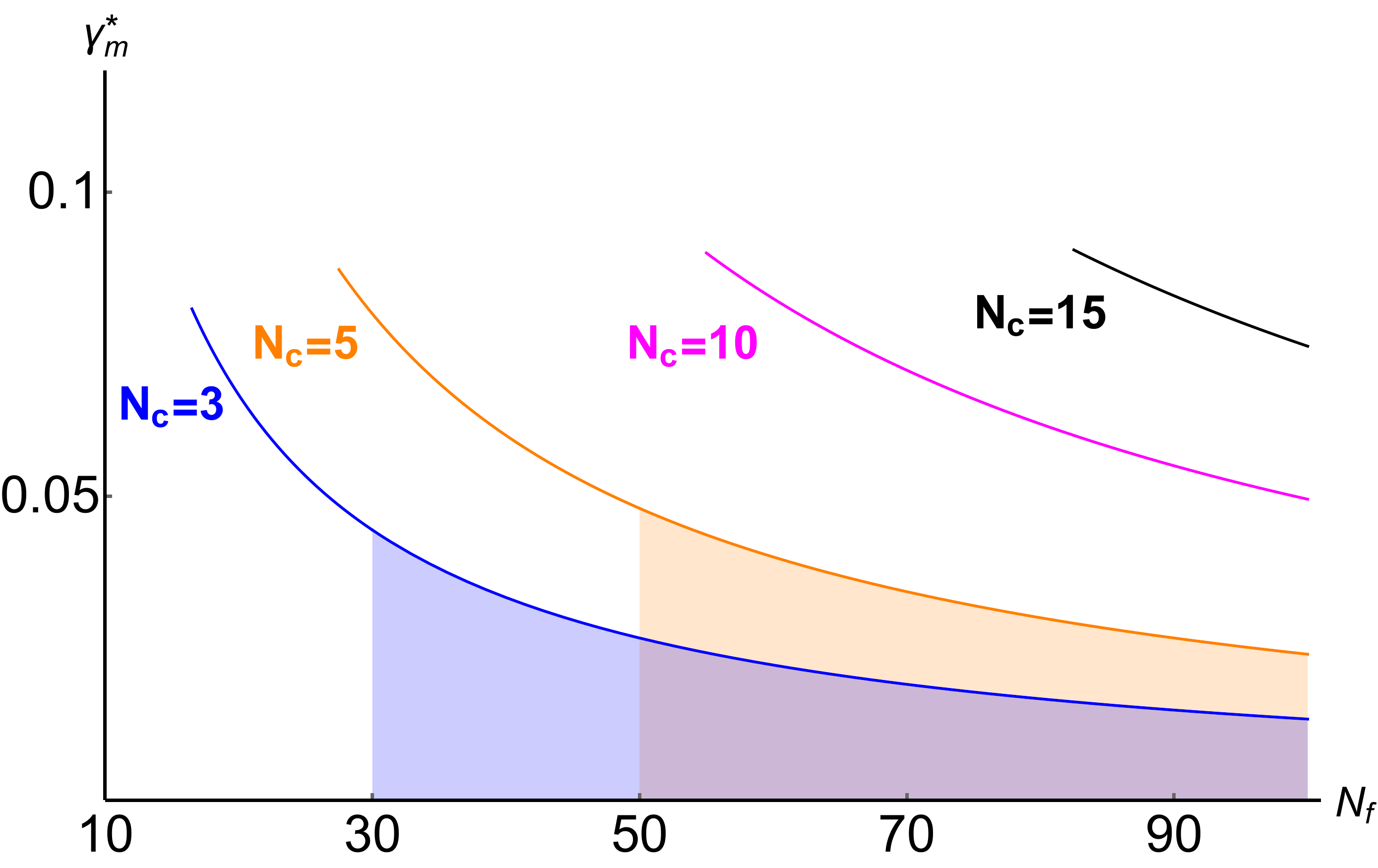} 
\caption{The anomalous dimension at the UVFP for $N_c=3,5,10,15$ (from left to right) for the fermions in the fundamental representation. The shaded regions start from $N_f=30(50)$ for $N_c=3(5)$ and correspond to the $1/N_f$  validity regions that for $N_c=10,15$ start for $N_f=100,150$ respectively.}
\label{gamma}
\end{figure}

Specialising to the case of the fundamental representation we have: 
\beq
\gamma_m^*(A) \stackrel{N_f\to \infty}\to   \frac{(N_c^2-1)}{2N_cN_f}\ .
\label{asymptoticfun}
\eeq
At relatively large $N_c$ this simplifies to  $\gamma_m^*(A)\to N_c/2 N_f$.
 Still with fermions in the fundamental representation we plot  in Fig.\ref{gamma} the anomalous dimension at the UVFP  as function of $N_f$ for distinct values of $N_c$. We nicely reproduce the results of Eq.~\ref{asymptoticfun} at large $N_f$.  
Requiring the first few known perturbative terms of the higher $1/N_f$ order expansion functions $H_i(A)$  ($i>1$) to be sufficiently small for $A$ as large as the radius of convergence ($A=3$) of $H_1(A)$, it was argued in \cite{Holdom:2010qs} that the expansion is applicable  for $N_c\lsim N_f/10$. This implies that only values of $\gamma_m^*(A)\lsim 1/20$ are acceptable. The regions where the $1/N_f$ expansion holds, for a fixed $N_c$, are shaded in Fig.\ref{gamma}.

The asymptotic behaviour in Eq.\ref{asymptotic}  holds also for the other matter representations. 
For example, for the adjoint representation we have $\exp \big[-a \frac{N_f}{N_c}+b\big]\ll$1 for any $N_f \geq 1$ and thus: 
\beq
\gamma_m^*(A) \approx \frac{C_R }{2 T_R N_f} =\frac{1}{2 N_f}\ .
\label{asymptoticADJ}
\eeq

Also, in contrast with the fundamental representation case, we find that for $N_f\gsim 7$ the $1/N_f$ expansion is trustable \emph{independently} of the value of $N_c$. The reason for this is that for the adjoint representation $C_G=C_R=T_R=N_c$ and therefore, up to the negligible $N_c$-dependence in the fourth-order group invariants appearing at the 4th-loop order in the beta function, the $N_c$ dependence in $H_{1,2,3,4}(A)$ cancels completely. For the $N_c$ dependence of the traditional conformal window we refer to \cite{Bergner:2015dya}.  This means that the large $N_f$ UVFPs will have $\gamma_m^*(A)\lsim 1/14$. A result remarkably close to $\gamma_m^*(A)\lsim 1/20$ for the fundamental representation.

We now confront our predictions for the safe anomalous mass dimensions with the bound coming from the conformal bootstrap. These derive from imposing crossing symmetry constraints on the 4-point function of a scalar (meson) operator $\Phi_{ij}$ transforming according to the bifundamental representation of the $SU(N_f)\times SU(N_f)$ global symmetry group. 
 From the  work of Nakayama \cite{Nakayama:2016knq} the bounds are $\gamma_m^*<1.79$ for $N_f=8$ and  $\gamma_m^*<1.88$ for $N_f=100$.  Clearly the values of the safe anomalous dimensions lie comfortably below this bound\footnote{We thank Nakayama for providing the $N_f = 100$ bootstrap value.}.

 \underline{\it Conformal window 2.0:}
We now use the information acquired above to delineate the complete, in $N_c$ and $N_f$, phase diagram for an $SU(N_c)$ gauge theory with fermionic matter in a given representation. We use as reference the line where  asymptotic freedom is lost, i.e.  $N_f^{AF}=11 C_G/(4T_R)$. As it is well known  decreasing $N_f$ slightly below this value  one achieves the perturbative Banks-Zaks infrared fixed point (IRFP), that at two loops yields $\alpha^*=-b_1/b_2$. This analysis has been extended to the maximum known order in perturbation theory \cite{Pica:2010xq,Ryttov:2010iz,Ryttov:2016ner}
and constitutes the state-of-the-art in this field. As we continue to lower the number of flavours, the IRFP becomes strongly coupled and at some critical $N_f^{IRFP}$,  is lost.  The lower boundary of the conformal window has been estimated analytically in different ways \cite{Sannino:2009za} 
 and tested via lattice simulations \cite{Pica:2017gcb}. 
 
\begin{figure*}[ht!]
\subfloat[Fundamental rep.]{\label{1a} \includegraphics[width=0.45\textwidth]{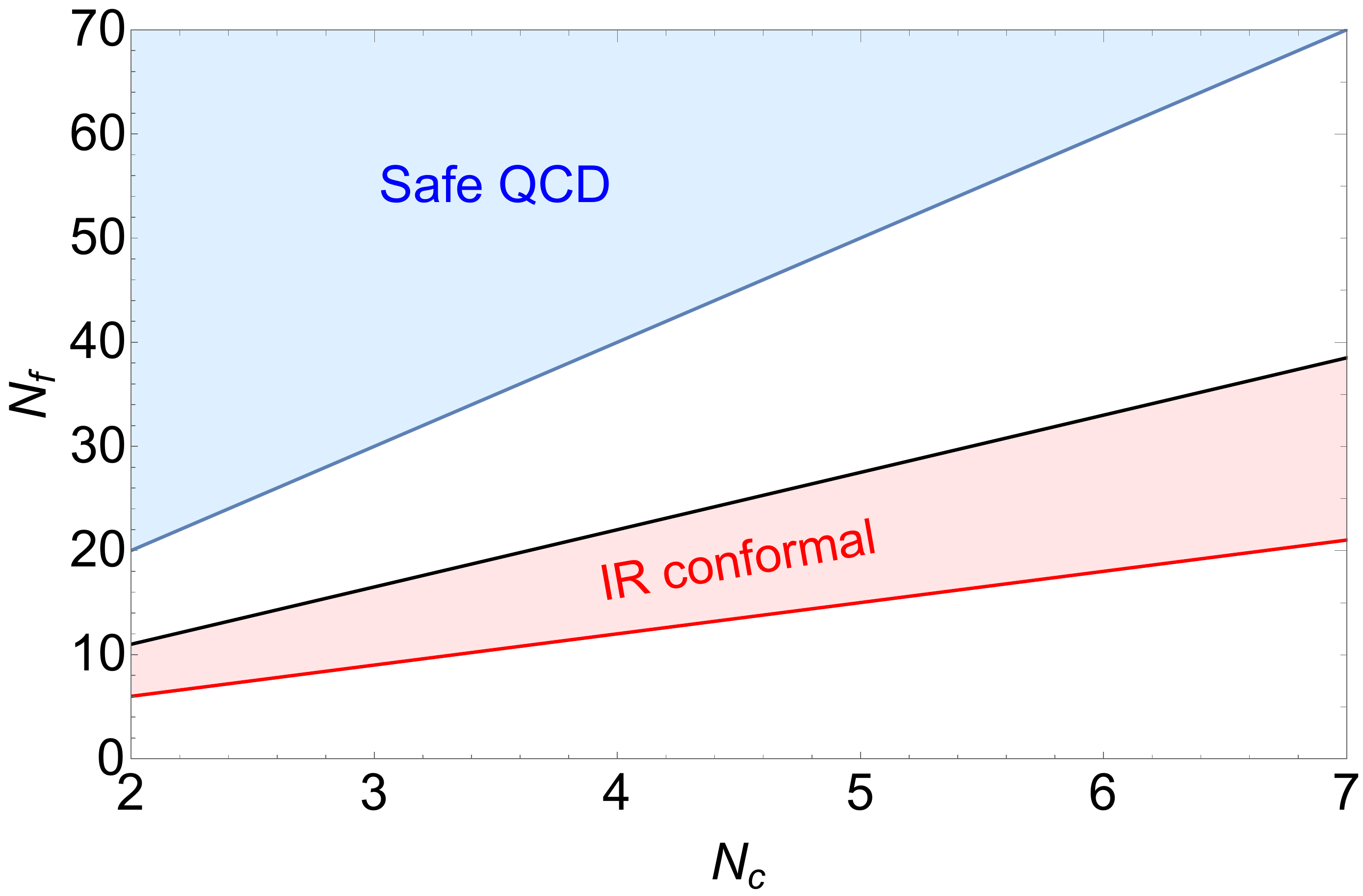}}
	\hfill
	\subfloat[Adjoint rep.]{\label{1b} \includegraphics[width=0.45\textwidth]{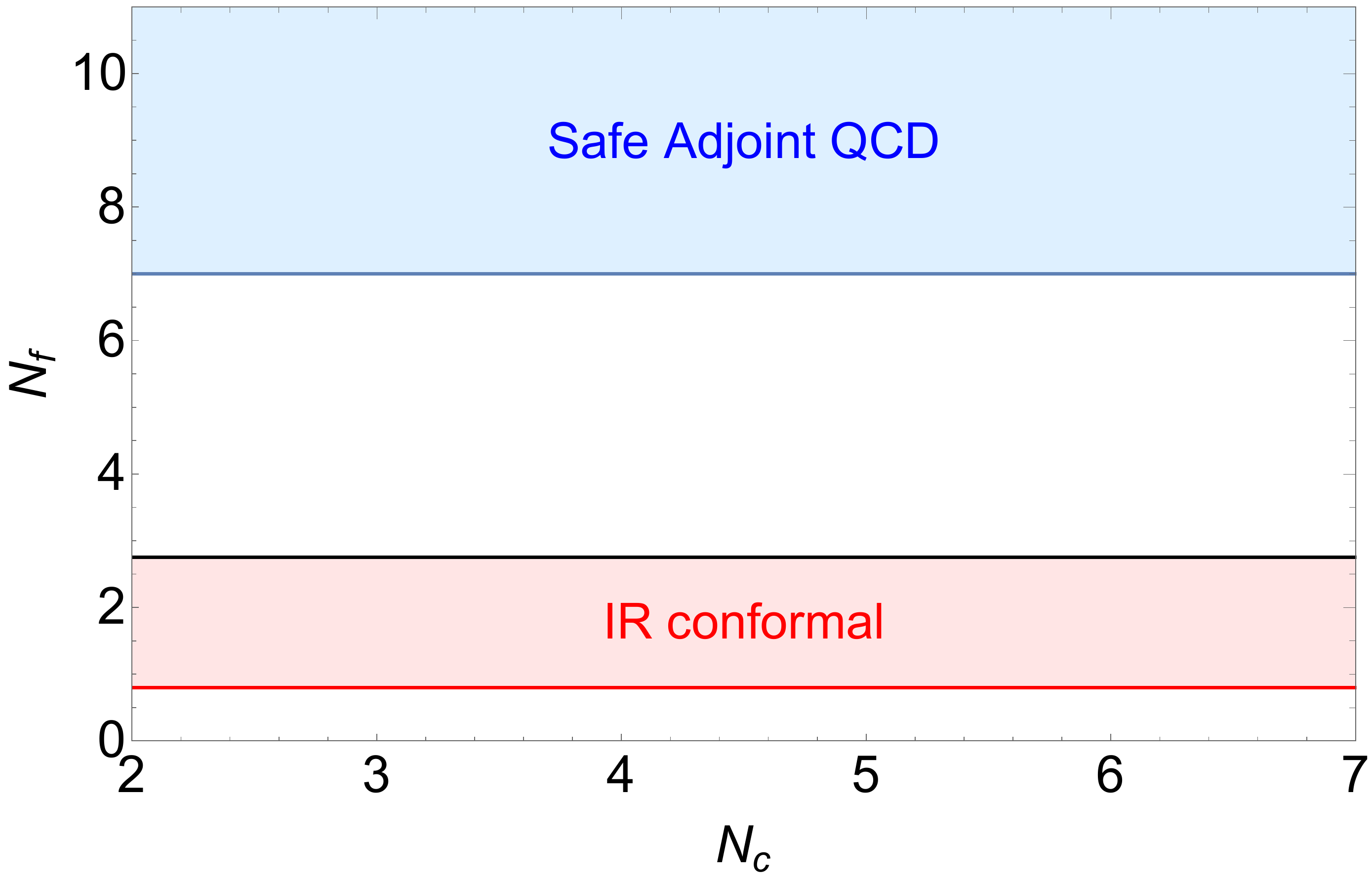}} 
\caption{Phase diagram of $SU(N_c)$ gauge theories with fermionic matter in the fundamental (left-panel) and adjoint (right-panel) representation. The shaded areas depict the corresponding conformal windows where the theories develop an IRFP (light red region) or an UVFP (light blue region).  The estimate of the lower boundary of the IRFP conformal window is taken from \cite{Pica:2010xq}. }
\label{CW}
\end{figure*}

 Just above the loss of asymptotic freedom, as mentioned in the introduction, Caswell \cite{Caswell:1974gg} demonstrated that no perturbative UVFP can emerge. By continuity there should be a region in colour-flavour space where the resulting theory is nonabelian QED with an unavoidable Landau pole. We will refer collectively to this region as {\it Unsafe QCD}. Here the theories can be viewed as low energy effective field theories with a trivial IRFP.  We then expect a critical value of number of flavours $N_f^{Safe}$ above which we achieve safety. This region extends to infinite values of $N_f$, i.e. the {\it Safe QCD} region. 
 Given that for the fundamental representation, the leading $1/N_f$ expansion is applicable only for $N_c\lsim N_f/10$ while for the adjoint representation we find $N_f\gsim 7$ for any $N_c$ it is sensible to use these as first estimate of the lower boundary of the {\it Safe QCD} region. Altogether, these constraints define the corresponding phase diagrams depicted in Fig.\ref{CW}. We conclude this discussion by commenting on the status of large $N_f$ super QCD. Using exact nonperturbative results it has been demonstrated that super QCD cannot be safe for any $N_f$ \cite{Intriligator:2015xxa}. 
 
 \underline{\it On abelian safety:}
Singularly interesting is the ultimate UV fate of abelian gauge theories. We investigate this by first rescaling the gauge coupling $A\equiv N_f\alpha/\pi$ in Eq.\ref{e10}.  This results in the $U(1)$ large $N_f$ $\beta$-function 
\begin{equation}
\frac{3}{2}\frac{\beta(A)}{A}=1+\sum_{i=1}^{\infty}\frac{F_i(A)}{N_f^i} \ , \quad F_1(A)=\int_0^{A/3} I_1(x) 
dx \ ,
\label{e2}\end{equation} 
with $I_1(x)$ the same as in the non-abelian case. 
Performing a Taylor expansion of the integrand in Eq.\ref{e2} and integrating term-by-term as for the non-abelian case Shrock  \cite{Shrock:2013cca} obtained the $nth$-loop coefficients $b_{n,n-1}$ with explicit results up to the 24th-loop order. Differently from the non-abelian case one finds, till the 24th order,  alternating signs for $b_{n,n-1}$ indicating a worse convergence for the abelian w.r.t. to the nonabelian case.  Nevertheless with this information we cannot yet exclude the possible existence of a stable UVFP. 
What we can, however, still determine at the would be fixed point is the correspondent fermion anomalous dimension. The latter is  related to the function $F_1(A)$ \cite{PalanquesMestre:1983zy,Holdom:2010qs} as follows: 
\begin{equation}
\gamma_m(A)= \frac{2A}{N_f}\frac{9}{(3-2A)(3+A)}\frac{dF_1(A)}{dA} +\mathcal{O}(1/N_f^2) \  .
\label{gamma1}
\end{equation}
Differently from the non-abelian case, the singularities in the $\gamma_m(A)$ and $\beta(A)$ happen at the same value of $A$ since the re-summation of the fermion bubbles is shared by both functions. Also, Eq.\ref{gamma1} relates the strength of the singularities with the logarithmic singularities in $F_1(A)$ manifested as simple poles in $G_1(A)$. 
The resulting UVFP to leading order in $1/N_f$ occurs at \cite{Holdom:2010qs}:
$A^*=\frac{15}{2}- 0.0117e^{-15\pi^2N_f/7}$.  
Inserting this value into the expression for the mass anomalous dimensions of Eq.\ref{gamma1} we obtain: 
\begin{equation}
\gamma_m^*(A^*)\approx\frac{e^{15\pi^2N_f/7}}{2\pi^2N_f\times 0.0117 } \ .
\end{equation}
The exponential proximity of the fixed point to the pole, generates the exponential growth in the number of flavours of the  mass anomalous dimension. For the physical case of $N_f\geq 1$, the corresponding $\gamma_m^*(A^*)$  exceeds the unitary bound that requires $\gamma_m(A^*)\leq 2$. This result suggests that the existence of an UVFP stemming from the resummation procedure for the abelian case  must be taken with the grain of salt and more work is needed to disentangle the ultimate ultraviolet fate of abelian gauge theories.

 Concluding, we briefly reviewed the salient  large $N_f$ results for nonabelian gauge-fermion theories.  These lead to the possible existence of an UVFP when asymptotic freedom is lost. To further test the emergence of {\it Safe QCD}-like theories we determined the related safe mass anomalous dimension.  We discovered that this important quantity is  controllably small. In particular for the fundamental representation we find that $\gamma_m^*(A)\lsim 1/20$ and for the adjoint case $\gamma_m^*(A)\lsim 1/14$. In fact  the safe anomalous dimension decreases with $N_f$ at finite $N_c$ for the fundamental, and independently of $N_c$ for the adjoint representation.  The so determined anomalous dimensions are comfortably within the current bootstrap bounds.  Our results lend support to the existence of two distinct regions in the colour-flavour plane when asymptotic freedom is lost.  The region contiguous to the loss of asymptotic freedom is {\it unsafe} with the theory being non-abelian QED in the IR and featuring an incurable Landau pole in the UV; and a second region starting above a new critical number of flavour line where safety is reached. The overall picture is summarised by the 2.0 upgraded version of the Conformal Window~\cite{Dietrich:2006cm} of Fig.~\ref{CW}.  
 For the $U(1)$ gauge theory the discovered exponential growth in the number of flavours of the safe mass anomalous dimension leaves unanswered the question of whether these theories can be safe at large number of flavours.

 The work of O.A. and F.S. are partially supported respectively by the Croatian Science Foundation (HRZZ) project "Terascale Physics for the LHC and Cosmos" as well as the H2020 CSA Twinning project No.692194, RBI-T-WINNING, and the Danish National Research Foundation grant DNRF:90.

\end{document}